\documentclass[12pt]{article}

\usepackage{axodraw}
\usepackage{epsf}
\usepackage{rotating}

\catcode`@=11
\def\citer{\@ifnextchar
[{\@tempswatrue\@citexr}{\@tempswafalse\@citexr[]}}

\def\@citexr[#1]#2{\if@filesw\immediate\write\@auxout{\string\citation{#2}}\fi
  \def\@citea{}\@cite{\@for\@citeb:=#2\do
    {\@citea\def\@citea{--\penalty\@m}\@ifundefined
       {b@\@citeb}{{\bf ?}\@warning
       {Citation `\@citeb' on page \thepage \space undefined}}%
\hbox{\csname b@\@citeb\endcsname}}}{#1}}
\catcode`@=12

\newcommand{\beq}{\begin{eqnarray}}
\newcommand{\eeq}{\end{eqnarray}}

\newcommand{\tgb}{\tan\beta}

\def\sla#1{\ifmmode%
\setbox0=\hbox{$#1$}%
\setbox1=\hbox to\wd0{\hss$/$\hss}\else%
\setbox0=\hbox{#1}%
\setbox1=\hbox to\wd0{\hss/\hss}\fi%
#1\hskip-\wd0\box1 } 

\newcommand{\lsim}{\raisebox{-0.13cm}{~\shortstack{$<$ \\[-0.07cm] $\sim$}}~}
\newcommand{\gsim}{\raisebox{-0.13cm}{~\shortstack{$>$ \\[-0.07cm] $\sim$}}~}
\newcommand{\fbi}{~fb$^{-1}\;$}
\begin{document}
 
\renewcommand{\thefootnote}{\fnsymbol{footnote} }
\begin{center}

{\large\sc Higgs Boson Production at Hadron Colliders: \\[0.3cm] Signal
and background Processes} \end{center}

\begin{center}
{David Rainwater$^1$, Michael Spira$^2$ and Dieter Zeppenfeld$^3$}

\vspace*{0.4cm}

{\it \small
$^1$ Fermilab, Batavia, IL, 60510, USA \\ 
$^2$ Paul Scherrer Institut, CH-5232 Villigen PSI, Switzerland \\
$^3$ Department of Physics, University of Wisconsin, Madison, WI 53706,
USA}
\end{center}
\vspace*{0.4cm}

\begin{abstract}
We review the theoretical status of signal and background calculations for
Higgs boson production at hadron colliders. Particular emphasis is given
to missing NLO results, which will play a crucial role for the Tevatron
and the LHC.
\end{abstract}

\section{Introduction}
The Higgs mechanism is a cornerstone of the Standard Model (SM) and its
supersymmetric extensions. Thus, the search for Higgs bosons is one of the
most important endeavors at future high-energy experiments.  In the SM
one Higgs doublet has to be introduced in order to break the
electroweak symmetry, leading to the existence of one elementary Higgs
boson, $H$ \cite{higgs}. The scalar sector of the SM is uniquely fixed by
the vacuum expectation value $v$ of the Higgs doublet and the mass $m_H$
of the physical Higgs boson \cite{hunter}.
The negative direct search for the Higgsstrahlung process $e^+e^-\to ZH$
at the LEP2 collider poses a lower bound of $114.1$ GeV on the SM
Higgs mass \cite{lep2}, while triviality arguments force the Higgs mass to be
smaller than $\sim 1$ TeV \cite{triviality}.

Since the minimal supersymmetric extension of the Standard Model (MSSM)
requires the introduction of two Higgs doublets in order to preserve
supersymmetry, there are five elementary Higgs particles, two CP-even
($h,H$), one CP-odd ($A$) and two charged ones ($H^\pm$). At lowest order
all couplings and masses of the MSSM Higgs sector are fixed by two
independent input parameters, which are generally chosen as
$\tgb=v_2/v_1$, the ratio of the two vacuum expectation values $v_{1,2}$,
and the pseudoscalar Higgs-boson mass $m_A$. At LO the light scalar
Higgs mass $m_h$ has to be smaller than the $Z$-boson mass $m_Z$.
Including the
one-loop and dominant two-loop corrections the upper bound is increased
to $m_h\lsim 135$ GeV \cite{mssmrad}. The negative direct searches for the
Higgsstrahlung processes $e^+e^-\to Zh,ZH$ and the associated production
$e^+e^-\to Ah,AH$ yield lower bounds of $m_{h,H} > 91.0$ GeV and
$m_A > 91.9$ GeV. The range $0.5 < \tgb < 2.4$ in the MSSM is
excluded by the Higgs searches at the LEP2 experiments \cite{lep2}.

The intermediate mass range, $m_H<196$~GeV at 95\% CL, is also favored
by a SM analysis of electroweak precision data~\cite{lep2}. In this 
contribution we will therefore concentrate on searches and measurements
for $m_H\lsim 200$~GeV. The Tevatron has a good chance to find 
evidence for such a Higgs boson, provided that sufficient integrated 
luminosity can be accumulated~\cite{run2}. The Higgs boson, if it exists, 
can certainly be seen at the LHC, and the LHC can provide measurements of the
Higgs boson mass at the $10^{-3}$ level~\cite{lhc}, and measurements of 
Higgs boson couplings at the 5 to 10\% level~\cite{Zeppenfeld:2000td}.
Both tasks, discovery and measurement of Higgs properties, require accurate
theoretical predictions of cross sections at the LHC, but these requirements
become particularly demanding for accurate coupling measurements.

In this contribution we review the present status of QCD calculations of signal
and background cross sections encountered in Higgs physics at hadron colliders.
Desired accuracy levels can be estimated by comparing to the statistical errors
in the determination of signal cross sections 
at the LHC. For processes like $H\to\gamma\gamma$, where
a very narrow mass peak will be observed, backgrounds can be accurately 
determined directly from data. For other decay channels, like $H\to b\bar b$
or $H\to\tau\tau$, mass resolutions of order 10\% require modest interpolation
from sidebands, for which reliable QCD calculations are needed. Most
demanding are channels like $H\to W^+W^-\to l^+l^-\sla p_T$, for which 
broad transverse mass peaks reduce Higgs observation to, essentially, a
counting experiment. Consequently, requirements on theory predictions vary
significantly between channels.
In the following we discuss production and decay channels in turn and
focus on theory requirements for the prediction of signal and background
cross sections. Because our main interest is in 
coupling measurements, we will not
consider diffractive channels in the following, which are
model-dependent and have large rate uncertainties~\cite{difractH}.
Potentially, they might contribute to Higgs 
discovery if, indeed, cross sections are sufficiently large.

\section{Gluon fusion}
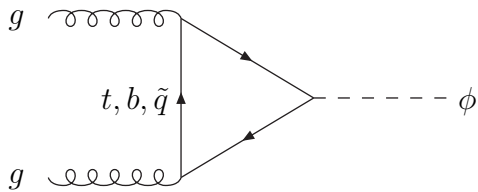
\begin{figure}[hbt]
\begin{center}
\setlength{\unitlength}{1pt}
\begin{picture}(180,100)(0,0)

\Gluon(0,20)(50,20){-3}{5}
\Gluon(0,80)(50,80){3}{5}
\ArrowLine(50,20)(50,80)
\ArrowLine(50,80)(100,50)
\ArrowLine(100,50)(50,20)
\DashLine(100,50)(150,50){5}
\put(155,46){$\phi$}
\put(20,46){$t,b,\tilde q$}
\put(-15,18){$g$}
\put(-15,78){$g$}

\end{picture}
\setlength{\unitlength}{1pt}
\caption[ ]{\label{fg:gghlodia} \it Typical diagram contributing to
$gg\to \phi$ at lowest order.}
\end{center}
\end{figure}

The gluon fusion mechanism $gg\to \phi$ provides the dominant production
mechanism of Higgs bosons at the LHC in the entire relevant mass range
up to about 1 TeV in the SM and for small and moderate values of $\tgb$
in the MSSM \cite{habil}. At the Tevatron this process plays the relevant role
for Higgs masses between about 130 GeV and about 190 GeV \cite{run2}. The gluon
fusion process is mediated by heavy quark triangle loops and, in the case
of supersymmetric theories, by squark loops in addition, if the squark
masses are smaller than about 400 GeV \cite{squark}, see 
Fig.~\ref{fg:gghlodia}.

\begin{figure}[hbt]
\hspace*{2.5cm} \includegraphics[width=10cm]{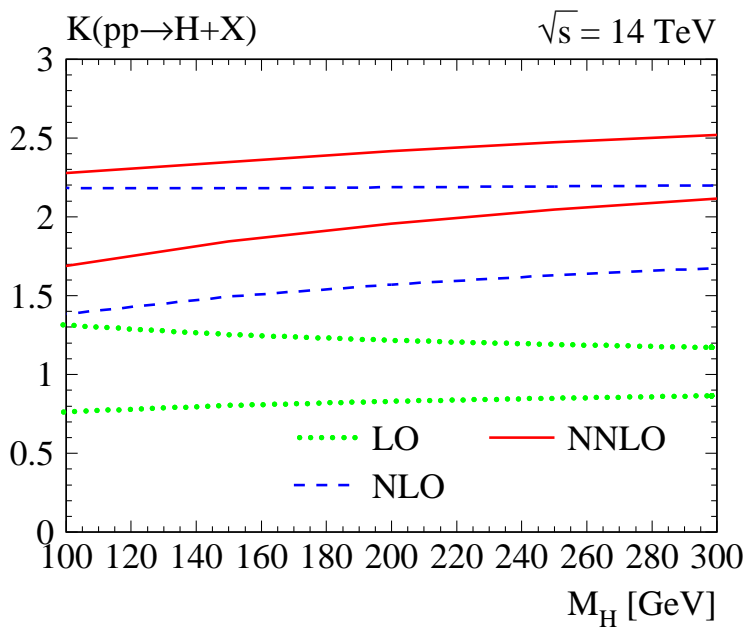}
      \caption[]{\it \label{fig:all14murf} Scale dependence of the $K$-factor 
      at the LHC. Lower curves for each pair are for 
      $\mu_R = 2m_H$, $\mu_F=m_H/2$, upper curves are for 
      $\mu_R =m_H/2$, $\mu_F=2m_H$.  The $K$-factor is
      computed with respect to the LO cross section at
      $\mu_R = \mu_F =m_H$. From Ref.~\cite{Harlander:2002wh}.}
\end{figure}

\begin{figure}
\hspace*{2cm} \includegraphics[width=8.5cm, angle=90]{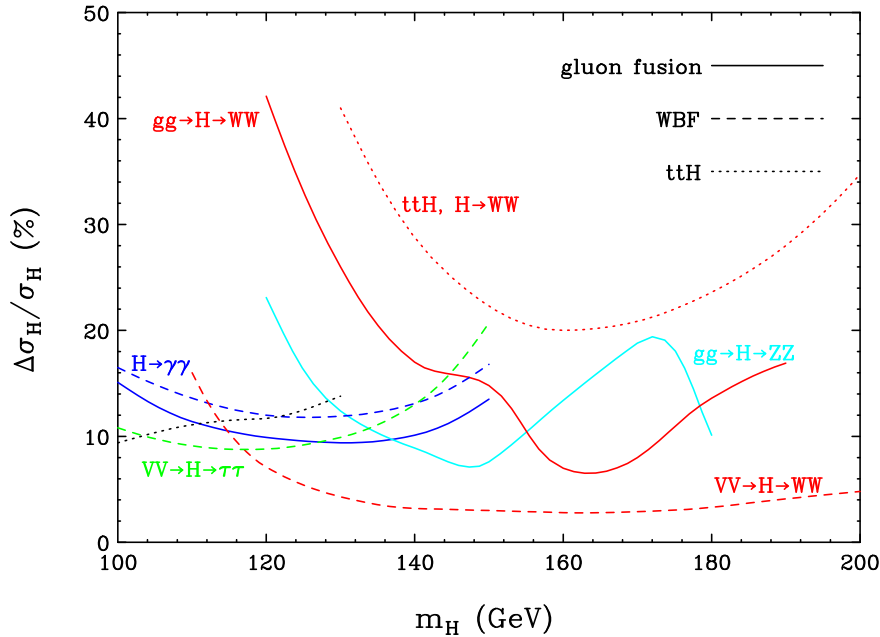}
\vspace*{-0.5cm}

 \caption{\it Expected relative error on the determination of 
$B\sigma$ for various Higgs search channels at the LHC 
with 200~\fbi of data~\cite{Zeppenfeld:2000td}. Solid 
lines are for inclusive Higgs production channels which 
are dominated by gluon fusion. Expectations for weak boson 
fusion are given by the dashed lines. Dotted lines are for 
$t\bar tH$ production with $H\to b\bar b$~\cite{drollinger:2001ym}
(black) and $H\to W^+W^-$~\cite{tth2ww}  (red). The latter assumes 
300\fbi of data.  }
 \label{fig:delsigh}
 \end{figure}

In the past the full two-loop QCD corrections have been determined. They
increase the production cross sections by 10--80\% \cite{glufusnlo},
thus leading to a significant change of the theoretical predictions. 
Very recently, Harlander and Kilgore have finished the full NNLO calculation, 
in the heavy top quark limit~\cite{3loop,Harlander:2002wh}.
This limit has been demonstrated to
approximate the full massive $K$ factor at NLO within 10\% for the SM Higgs
boson in the entire mass range up to 1 TeV \cite{softgluon}. Thus, a similar
situation can
be expected at NNLO. The reason for the quality of this approximation is
that the QCD corrections to the gluon fusion mechanism are dominated by
soft gluon effects, which do not resolve the one-loop Higgs coupling to gluons.
Fig.~\ref{fig:all14murf} shows the resulting $K$-factors at the LHC and
the scale variation of the $K$-factor. The calculation stabilizes at NNLO,
with remaining scale variations at the 10 to 15\% level. 
These uncertainties are comparable to the experimental errors which can 
be achieved with 200\fbi of data at the LHC, see solid lines in 
Fig.~\ref{fig:delsigh}. 
The full NNLO results confirm earlier estimates which were obtained 
in the frame work of soft gluon resummation
\cite{softgluon} and soft approximations \cite{soft} of the full
three-loop result. The full soft gluon resummation has been performed in
Ref.\,\cite{soft2}. The resummation effects enhance the NNLO result by
about 10\% thus signaling a perturbative stabilization of the
theoretical prediction for the gluon-fusion cross section.

In supersymmetric theories the gluon fusion cross sections for
the heavy Higgs, $H$, and, for small $m_A$, also for the light Higgs, $h$, 
may be dominated by
bottom quark loops for large values of $\tgb\gsim 10$ so that the heavy
top quark limit is not applicable. This can be clearly seen in the NLO
results, which show a decrease of the $K$ factor down to about 1.1 for
large $\tgb$ \cite{glufusnlo}. This decrease originates from an interplay
between the large
positive soft gluon effects and large negative double logarithms of the
ratio between the Higgs and bottom masses. 
In addition, the shape of the $p_T$ distribution of the Higgs boson may be 
altered; if the bottom loop is dominant, the $p_T$
spectrum becomes softer than in the case of top-loop dominance. These effects
lead to some model dependence of predicted cross sections.

%

Let us now turn to a discussion of backgrounds for individual decay modes.
\vspace{0.1in}

\noindent
\underline{\it (i) $\phi\to \gamma\gamma$} \\
At the LHC the SM Higgs boson can be found in the mass range up to 
about 150~GeV by means of the rare photonic decay 
mode $H\to\gamma\gamma$~\cite{lhc}. The dominant Higgs decays 
$H\to b\bar b, \tau^+\tau^-$ are overwhelmed by large 
QCD backgrounds in inclusive searches. 
The QCD $\gamma\gamma$ background is known at NLO,
including all relevant fragmentation effects. The present status is
contained in the program DIPHOX \cite{diphox}. 
The loop mediated process $gg\to\gamma\gamma$ contributes about 50\% to
the $\gamma\gamma$ background and has been calculated at NLO very
recently \cite{gggamgam}. However, a numerical analysis of the two-loop
result is still missing.

Once the experiment is performed, the diphoton background can be determined
precisely from the data, by a measurement of $d\sigma/dm_{\gamma\gamma}$
on both sides of the resonance peak. The NLO calculations are useful, 
nevertheless, for an accurate prediction of expected accuracies and for a
quantitative understanding of detector performance.
\vspace{0.1in}

\noindent
\underline{\it (ii) $H\to W^+W^-$} \\
This mode is very important for Higgs masses above $W$-pair but below
$Z$-pair threshold, where $B(H\to WW)$ is close to 100\%.
In order to suppress the $t\bar t\to b\bar bW^+W^-$ background for 
$W^+W^-$ final states, a jet veto is crucial. However, gluon fusion 
receives sizeable contributions from real gluon bremsstrahlung at NLO,
which will also be affected by the jet veto. These effects have recently
been analyzed in Ref.~\cite{Catani:2001cr}, in the soft approximation 
to the full NNLO calculation. A veto of additional jets 
with $p_{Tj}>15$~GeV, as e.g. envisioned by ATLAS~\cite{lhc}, 
reduces the NNLO $K$-factor
to about $K=0.8$\footnote{It should be noted that for this strong cut in
$p_{Tj}$ the NNLO result may be plagued by large logarithms of this cut,
which have to be resummed, see \cite{soft2}.}, i.e. one loses more than
60\% of signal events. 
In addition the scale dependence of the cross section starts to grow with
such stringent veto criteria. These effects need to be modeled with a
NLO Monte Carlo program for $H+jet$ production in order to reach a 
reliable quantitative result for the signal rate. Since stop and sbottom 
loops are sizeable in supersymmetric theories for squark
masses below about 400 GeV, their inclusion is important in these
investigations.

From the perspective of background calculations, $H\to WW$ is the most 
challenging channel. Backgrounds are of the order of the signal rate or larger,
which requires a 5\% determination or better for the dominant background cross
sections in order to match the statistical power of LHC experiments. In
fact, the large errors at $m_H\lsim 150$~GeV depicted in Fig.~\ref{fig:delsigh}
($gg\to H\to WW$ curve) are dominated by an assumed 5\% background 
uncertainty. Clearly, such small errors cannot be achieved by NLO calculations
alone, but require input from LHC data. 
Because of two missing neutrinos in the 
$W^+W^-\to l^+l^-\sla p_T$ final state, the Higgs mass cannot be reconstructed
directly. Rather, only wide $(l^+l^-;\sla p_T)$ transverse mass distributions
can be measured, which do not permit straightforward sideband 
measurements of the 
backgrounds. Instead one needs to measure the normalization of the backgrounds
in signal poor regions and then extrapolate these, with the help of 
differential cross sections predicted in perturbative QCD, to the signal
region. The theory problem is the uncertainty in the shape of the
distributions used for the extrapolation, which will depend on an 
appropriate choice of the ``signal poor region''. No analysis of the 
concomitant uncertainties, at LO or NLO QCD, is available to date.

After the jet veto discussed above, 
the dominant background processes are $pp\to W^+W^-$ and
(off-shell) $t\bar t$ production~\cite{lhc}. $W^+W^-$ production is known 
at NLO~\cite{vvnlo} and available in terms of parton level Monte Carlo
programs. In addition, a full NLO calculation including spin correlations
of the leptonic $W,Z$ decays, in the narrow width approximation, 
is available \cite{kunszt}. For Higgs boson
masses below the $W^+W^-(ZZ)$ threshold, decays into $WW^*(ZZ^*)$ are
important \cite{offshell,habil}. Since hadron colliders will be sensitive
to these off-shell
tails, too, the backgrounds from $VV^*$ production become relevant. There
is no NLO calculation of $VV^*$ background processes available so far, so
that it is not clear if NLO effects will be significant in the tails of
distributions needed for the Higgs search in these cases. Moreover, for
$WW^*$ production the inclusion of spin correlations among the final state
leptons is mandatory \cite{dittdrei}. 

Top quark backgrounds arise from top-pair and $tWb$ production.
Recently, a new theoretical analysis of $pp\to t^{(*)}\bar t^{(*)}$ has
become available including full lepton correlations and off-shell effects 
of the final state top quarks arising from the non-zero top decay 
width~\cite{ttoff}. This calculation automatically
includes $pp\to tbW$ and those contributions to $pp\to b\bar b W^+W^-$,
which are gauge-related to $tbW$ couplings and describes
the relevant tails for the Higgs search at LO. It is now necessary to
investigate the theoretical uncertainties of this background.
A NLO calculation of off-shell top-pair production may well be needed
to reach the required 5\% accuracy for extrapolation to the Higgs search
region.

Other important reducible backgrounds are the $Wt\bar t,Zt\bar t,Wb\bar b$ 
and $Zb\bar b$ production processes. While $Vt\bar t$ ($V=W,Z$) production 
is only known
at LO, the associated vector boson production with $b\bar b$ pairs is
known at NLO including a soft gluon resummation \cite{veseli}. Thus $Vb\bar b$
production can be considered as reliable from the theoretical point of
view, while a full NLO calculation for $Vt\bar t$ production is highly
desirable, since top mass effects will play a significant role. In
addition, the background from $gb \to tH^-, g\bar b\to \bar t H^+$ has to
be taken into account within the MSSM framework. The full LO matrix 
elements are included in the
ISAJET Monte Carlo program, which can easily be used for experimental
analyses.

\vspace{0.1in}
\noindent
\underline{\it (iii) $H\to ZZ \to 4\ell^\pm$} \\
A sharp Higgs peak can be observed in the four lepton invariant mass 
distribution. Hence, the $ZZ\to 4\ell^\pm$ backgrounds are directly 
measurable in the sidebands and can safely be interpolated to the 
signal region. 

\section{$qq\to qqH$}
In the SM the $WW,ZZ$ fusion processes $qq\to qqV^*V^*\to qqH$ play a
significant role at the LHC for the entire Higgs mass range up to 1 TeV. 
We refer to them as weak boson fusion (WBF). The WBF
cross section becomes comparable to the gluon fusion cross
section for Higgs masses beyond $\sim 600$ GeV \cite{habil} and is sizable,
of order 20\% of $\sigma(gg\to H)$, also in the intermediate mass region. 
The energetic quark jets in the
forward and backward directions allow for additional cuts to suppress the
background processes to WBF. The NLO QCD corrections
can be expressed in terms of the conventional corrections to the DIS
structure functions, since there is no color exchange between the two
quark lines at LO and NLO. NLO corrections increase the production cross 
section by about 10\% and are thus small and under theoretical 
control~\cite{vvhnlo,susyqcd}. These small theory uncertainties make WBF a very 
promising tool for precise coupling measurements. However, additional 
studies are needed to assess the
theoretical uncertainties associated with a central jet veto. This veto
enhances the color singlet exchange of the signal over color octet 
exchange QCD backgrounds~\citer{wbfgamma,wbf.wwlomh}. 

In the MSSM, first parton level
analyses show that it should be possible to cover the
full MSSM parameter range by looking for the light Higgs decay 
$h\to \tau^+\tau^-$ (for $m_A\gsim 150$~GeV) and/or the heavy Higgs 
$H\to\tau^+\tau^-$resonance (for a relatively small $m_A$) 
in the vector-boson fusion processes~\cite{vvhtau}.
Although these two production processes are suppressed with respect
to the SM cross section, their sum is of SM strength.

For the extraction of Higgs couplings it is important to distinguish 
between WBF and gluon fusion processes which lead to $H+jj$ final states.
With typical WBF cuts, including a central jet veto, gluon fusion 
contributions are expected at order 10\% of the WBF cross section, i.e. 
the contamination is modest~\cite{hjjmass,wbf.exp}.
The gluon fusion processes are mediated by heavy top and
bottom quark loops, in analogy to the LO gluon fusion diagram of 
Fig.~\ref{fg:gghlodia}. The full massive cross section for $H+jj$ 
production via gluon fusion has been obtained only recently~\cite{hjjmass},
while former analyses were performed in the heavy top quark 
limit~\cite{hjjlimit}. Since stop and sbottom loops yield a sizeable 
contribution to the inclusive gluon fusion cross section, a similar 
feature is expected for $H+jj$
production. Thus, it is important to compute the effects of stop and
sbottom loops in $H+jj$ gluon fusion processes, which has not been done 
so far. 

\vspace{0.1in}
\noindent
\underline{\it (i) $H\to\gamma\gamma$} \\
Parton level analyses show that $H\to \gamma\gamma$ decays in WBF Higgs
production can be isolated with signal to background ratios of order 
one~\cite{wbfgamma} and with statistical errors of about 15\%, for 200\fbi of 
data (see Fig.~\ref{fig:delsigh}). Like for the inclusive $H\to\gamma\gamma$ 
search, background levels can be precisely determined from a sideband analysis
of the data. Prior to data taking, however, full detector simulations are 
needed to confirm the parton level results and improve on the search 
strategies. 

\begin{figure}[hbt]
\begin{center}
\setlength{\unitlength}{1pt}
\begin{picture}(200,170)(5,-30)

\Gluon(0,20)(50,20){-3}{5}
\Gluon(0,80)(50,80){3}{5}
\Gluon(90,-5)(140,-35){-3}{5}
\Gluon(90,105)(140,135){3}{5}
\Photon(130,80)(180,80){3}{5}
\Photon(130,20)(180,20){3}{5}
\ArrowLine(50,20)(50,80)
\ArrowLine(50,80)(90,105)
\ArrowLine(90,105)(130,80)
\ArrowLine(130,80)(130,20)
\ArrowLine(130,20)(90,-5)
\ArrowLine(90,-5)(50,20)
\put(55,46){$q$}
\put(-15,18){$g$}
\put(-15,78){$g$}
\put(145,133){$g$}
\put(145,-37){$g$}
\put(185,78){$\gamma$}
\put(185,18){$\gamma$}
\put(215,48){$+$}

\end{picture}
\begin{picture}(180,180)(-45,-30)

\Gluon(0,20)(50,20){-3}{5}
\Gluon(0,80)(50,80){3}{5}
\Gluon(25,20)(75,-10){-3}{5}
\Gluon(25,80)(75,120){3}{5}
\Photon(100,80)(150,80){3}{5}
\Photon(100,20)(150,20){3}{5}
\ArrowLine(50,20)(50,80)
\ArrowLine(50,80)(100,80)
\ArrowLine(100,80)(100,20)
\ArrowLine(100,20)(50,20)
\put(55,46){$q$}
\put(-15,18){$g$}
\put(-15,78){$g$}
\put(80,118){$g$}
\put(80,-12){$g$}
\put(155,78){$\gamma$}
\put(155,18){$\gamma$}

\end{picture}
\setlength{\unitlength}{1pt}
\caption[ ]{\label{fg:gamgamjj} \it Typical diagrams contributing to
$gg\to \gamma\gamma jj$ at lowest order.}
\end{center}
\end{figure}
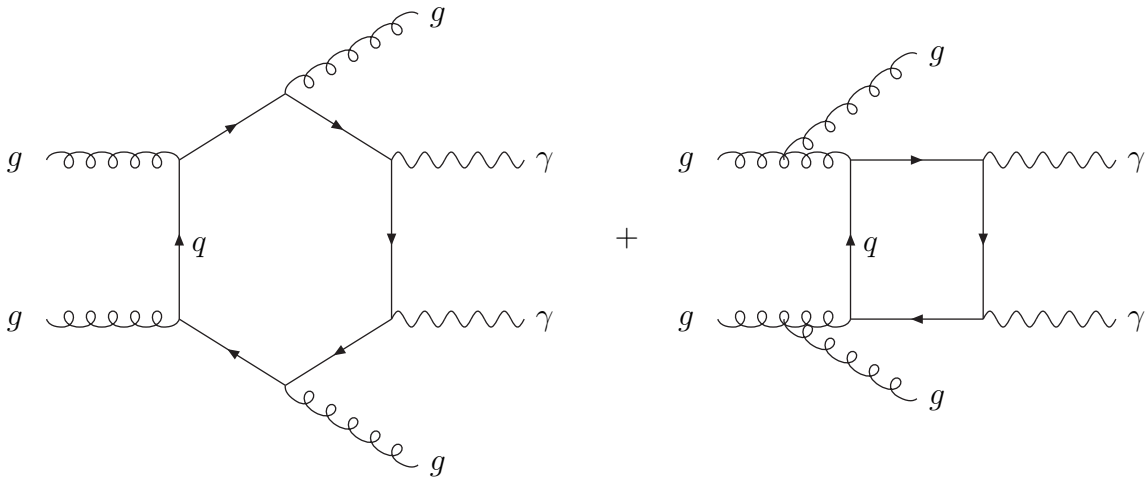

Improved background calculations are desirable as well. In particular,
the $pp\to \gamma \gamma jj$ background via quark loops (see
Fig.~\ref{fg:gamgamjj}) has not been calculated so far.

\noindent
\underline{\it (ii) $H\to\tau^+ \tau^-$} \\
The observation of $H\to\tau\tau$ decays in WBF will provide crucial 
information on Higgs couplings to fermions~\cite{Zeppenfeld:2000td}
and this channel alone guarantees Higgs observation within the 
MSSM~\cite{vvhtau} and may be an important discovery channel
at low pseudoscalar mass, $m_A$. Recent detector 
simulations~\cite{wbf.exp} confirm parton level results~\cite{wbftau} for 
the observability of this channel. (See Fig.~\ref{fig:delsigh} for parton 
level estimates of statistical errors.)

\begin{figure}[thb]
\hspace*{2cm} \includegraphics[width=9.0cm, angle=90]{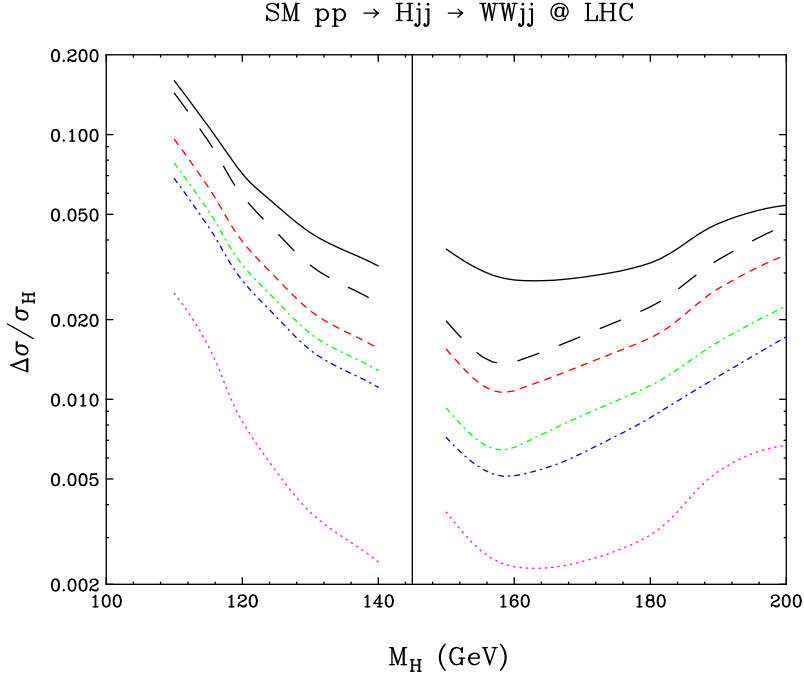}%
 \caption{\it Contributions of background systematic errors $\Delta\sigma$ 
to a measurement of $\sigma_H=\sigma B(H\to WW)$ in WBF. 
Shown, from bottom to top, are the effects of 
a 10\% uncertainty of the $\tau\tau jj$ rate (dotted line),
a 50\% error on the QCD WWjj rate (blue dash-dotted),
a 30\% error on the electroweak WWjj rate (green dash-dotted), and 
a 10\% error on $\sigma(t\bar t+$jets) (red dashes). 
The long-dashed line adds these errors in quadrature. 
For comparison, the solid line shows the expected statistical 
error for 200\fbi.
The vertical line at 145 GeV separates analyses optimized 
for small~\cite{wbf.wwlomh}  and large~\cite{wwhimH} Higgs masses.
} \label{fig:errorbugetWW}
 \end{figure}

The $\tau^+ \tau^-$-invariant mass can be reconstructed at the LHC with
a resolution of order 10\%. This is possible in the $qq\to qqH$ mode 
because of the large transverse momentum of the Higgs. 
In turn this means a sideband analysis can be used,
in principle, to directly measure backgrounds. The most important of
these backgrounds is QCD $Zjj$ production (from QCD corrections to Drell-Yan)
or electroweak $Zjj$ production via WBF~\cite{wbftau}. The (virtual)
Z (or photon) then decays into a $\tau^+\tau^-$ pair. These $Zjj$ backgrounds,
with their highly nontrivial shape around $m_{\tau\tau}\approx m_Z$, can be 
precisely determined be observing $Z\to e^+e^-,\mu^+\mu^-$ events in identical
phase space regions. Theoretically the QCD $Zjj$ background is under control
also, after the recent calculation of the full NLO corrections~\cite{zjjNLO}.
For the $\tau^+\tau^-$ backgrounds the inclusion of $\tau$ polarization
effects is important in order to obtain reliable tau-decay distributions
which discriminate between signal processes
($h,H\to \tau^+\tau^-$) and backgrounds. This can be achieved by linking
the TAUOLA program \cite{tauola} to existing Monte Carlo programs.

\vspace{0.1in}
\noindent
\underline{\it (iii) $H\to WW\to \ell^+ \ell^- \sla p_T$} \\
The most challenging WBF channel is $H\to WW^{(*)}$ decay which does not
allow for direct Higgs mass reconstruction and, hence, precludes a simple
sideband determination of backgrounds. The important 
backgrounds~\cite{wwhimH,wbf.wwlomh} involve
(virtual) $W$ pairs, namely top decays 
in $t\bar t+$jets production, and QCD and electroweak $WWjj$ production.
QCD and EW $\tau\tau jj$ production are subdominant after cuts, they are
known at NLO~\cite{zjjNLO}, and they can be determined directly, in phase 
space regions for jets which are identical to the signal region and with
high statistics, by studying
$e^+e^-$ or $\mu^+\mu^-$ pairs instead of $\tau^+\tau^-$. 

Demands on QCD
calculations can be estimated by comparing the effects of systematic 
background errors on the measurement of the signal rate with statistical 
errors achievable at the LHC with 200\fbi of data. Results are shown in
Fig.~\ref{fig:errorbugetWW} for an assumed 10\% error on 
$\sigma(t\bar t+$jets), a 50\% error on the QCD WWjj rate, and
a 30\% error on the electroweak WWjj rate. The latter two should be achievable
from a LO extrapolation from signal poor to signal rich regions of phase space.
A 10\% error of $\sigma(t\bar t+$jets), on the other hand, may require a NLO
calculation, in particular of the on-shell $t\bar t+1$~jet cross section
which dominates the $t\bar t$ background. 
Off-shell effects have recently been studied at 
LO~\cite{ttoff} and a ${\cal O}(20\%)$ increase of the $t\bar t$ background 
was found, which, presumably,
is small enough to permit the inclusion of off-shell effects at LO only.
However, a dedicated study is needed to devise optimal techniques for a 
reliable background determination for $H\to WW$ searches in WBF, for all major 
backgrounds.

\vspace{0.1in}
\noindent
\underline{\it (iv) Jet veto and Jet Tagging} \\
Background suppression in the WBF channels relies on double forward jet 
tagging to identify the scattered quark jets of the $qq\to qqH$ signal
and it employs a veto of relatively soft central jets 
(typically of $p_T>20$~GeV) to exploit the different gluon radiation patterns
and QCD scales of $t$-channel color singlet versus color octet exchange. 
Transverse momenta of these tagging or veto jets are relatively small for 
fixed order perturbative calculations of hard processes at the LHC. Thus,
dedicated studies will be needed to assess the applicability of NLO QCD for 
the modeling of tagging jets in WBF and for the efficiency of a central jet 
veto in the Higgs signal. First such studies have been performed in the past 
at LO, for $Wjj$ or $Zjj$ events~\cite{Chehime:1992ub}. While NLO Monte Carlos
for QCD $Vjj$ production are now available~\cite{zjjNLO,MCFM}, 
the corresponding
NLO determination of electroweak $Vjj$ cross sections would be highly 
desirable. This would allow a comparison of calculated and measured 
veto efficiencies in a WBF process. These efficiencies must be known at the 
few percent level for the signal in order to extract Higgs couplings without
loss of precision.

At present, the veto efficiencies for signal and background processes are 
the most uncertain aspect of WBF Higgs production at the LHC. Any improvement
in their understanding, from QCD calculations, from improved Monte Carlo tools,
or from hadron collider data would be very valuable.

\section{$t\bar t\phi$ production}
SM Higgs boson production in association with $t\bar t$ pairs plays a
significant role at the LHC for Higgs masses below about 130 GeV, since
this production mechanism makes the observation of 
$H\to b\bar b$ possible \cite{lhc,drollinger:2001ym,tth2bb.atlas,tth2bb.cms}. 
The decay $H\to\gamma\gamma$ is 
potentially visible in this channel at high integrated luminosity. For Higgs 
masses above about 130 GeV, the decay $H\to W^+W^-$ can be 
observed~\cite{tth2ww}. 
$t\bar tH$ production could conceivably be used to determine the top Yukawa 
coupling directly from the cross section, but this requires either assumptions 
about the branching ratio for $H\to b\bar{b}$, which are not justified in 
extensions of the SM, or observability of decay to either $\gamma\gamma$
or $W^+W^-$. 
Recently, the
NLO QCD corrections have become available. They decrease the cross section
at the Tevatron by about 20--30\% \cite{tthnlo,tthnloq}, while they
increase the signal rate at the LHC by about 20--40\% \cite{tthnlo}. The
scale dependence of the production cross
section is significantly reduced, to a level of about 15\%, which can be
considered as an estimate of the theoretical uncertainty. Thus, the signal
rate is under proper theoretical control now. In the MSSM, $t\bar th$
production with $h\to \gamma\gamma,b\bar b$ is important at the LHC in
the decoupling regime, where the light scalar $h$ behaves as the SM
Higgs boson \cite{lhc,drollinger:2001ym,tth2bb.atlas,tth2bb.cms}. 
Thus, the SM results can also be used for $t\bar{t}h$ 
production in this regime. \\

\noindent
\underline{\it (i) $t\bar t\phi\to t\bar t b\bar b$} \\
The major backgrounds to the $\phi\to b\bar b$ signal in associated $t\bar
t\phi$ production come from $t\bar t jj$ and $t\bar t b\bar b$ production,
where in the first case the jets may be misidentified as $b$ jets. A full
LO calculation is available for these backgrounds and will be included in
the conventional Monte Carlo programs. However, an analysis of the
theoretical uncertainties is still missing. A first step can be made by
studying the scale dependence at LO in order to investigate the effects on
the total normalization and the event shapes. But for a more sophisticated
picture a full NLO calculation is highly desirable. A second question is
whether these backgrounds can be measured in the experiments off the Higgs
resonance and extrapolated to the signal region. \\

\noindent
\underline{\it (ii) $t\bar t\phi\to t\bar t \gamma\gamma$} \\
The $t\bar t \gamma\gamma$ final states develop a narrow resonance in the
invariant $\gamma\gamma$ mass distribution, which enables a measurement
of the $t\bar t \gamma\gamma$ background directly from the sidebands.\\

\noindent
\underline{\it (iii) $t\bar t\phi\to t\bar t W^+W^-$} \\
This channel does not allow reconstruction of the Higgs. Instead, it relies 
on a counting experiment of multiplepton final states where the background 
is of approximately the same size as the signal. The principal backgrounds are 
$t\bar{t}Wjj$ and $t\bar{t}\ell^+\ell^- (jj)$, with minor backgrounds of 
$t\bar{t}W^+W^-$ and $t\bar{t}t\bar{t}$. For the $3\ell$ channel, the 
largest background is $t\bar{t}\ell^+\ell^-$ where one lepton is lost. 
It is possible that this rate could be measured directly for the lepton pair 
at the $Z$ pole and the result extrapolated to the signal region of phase 
space. However, for $t\bar{t}Vjj$ backgrounds the QCD uncertainties 
become large and unknown, due to the presence of two additional soft jets 
in the event. Further investigation of these backgrounds is essential, and 
will probably require comparison with data, which is not expected to be 
trivial.

\section{$b\bar b\phi$ production}
In supersymmetric theories $b\bar b\phi$ production becomes the dominant
Higgs boson production mechanism for large values of $\tgb$ \cite{habil},
where the
bottom Yukawa coupling is strongly enhanced. In contrast to $t\bar t\phi$
production, however, this process develops potentially large logarithms,
$\log m_\phi^2/m_b^2$, in the high-energy limit due to the smallness of the
bottom quark mass, which are
related to the development of $b$ densities in the initial state. They can
be resummed by evolving the $b$ densities according to the
Altarelli--Parisi equations and introducing them in the production
process \cite{willenbrock}. The introduction of conventional $b$ densities
requires an
approximation of the kinematics of the hard process, i.e.~the initial $b$
quarks are assumed to be massless, have negligible transverse momentum
and travel
predominantly in forward and backward direction. These approximations can be
tested in the full $gg\to b\bar b\phi$ process. At the Tevatron it turns
out that they are not valid so that the effective cross
section for $b\bar b\to \phi$ has to be considered as an overestimate of
the resummed result. An improvement of this resummation requires an
approach which describes the kinematics of the hard process in a
better way. Moreover, since the experimental analyses require 3 or 4 $b$
tags \cite{run2,lhc}, the spectator $b$ quarks need to have a sizeable
transverse momentum of at least 15--20 GeV. Thus a resummation of a
different type of potentially arising logarithms, namely
$\log m_\phi^2/(m_b^2+p_{tmin}^2)$ is necessary. This
can be achieved by the introduction of e.g.~unintegrated parton
densities \cite{uninpdf} or an extension of the available resummation
techniques.

As a first step, however, we have to investigate if the energy of the
Tevatron and LHC is sufficiently large to develop the factorization of
bottom densities. This factorization requires that the transverse
momentum distribution of the (anti)bottom quark scales like
$d\sigma/dp_{Tb}\propto p_{Tb}/(m_b^2+p_{Tb}^2)$ for transverse momenta
up to the factorization scale of the (anti)bottom density.
\begin{figure}[hbt]
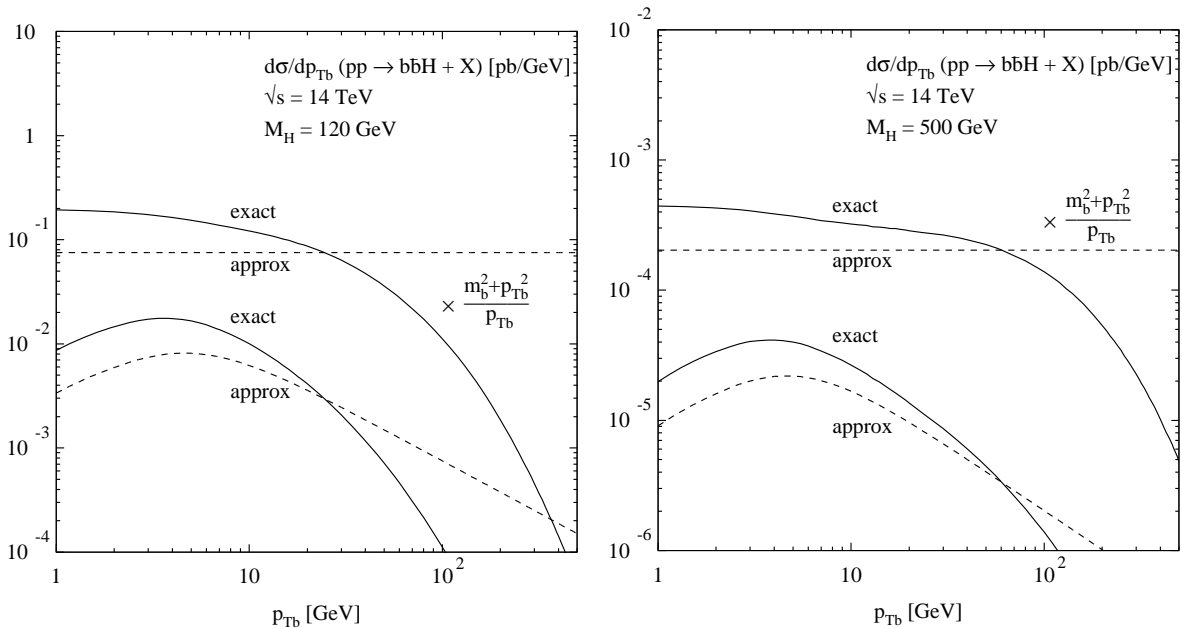

\vspace*{-0.5cm}
\hspace*{0.5cm}
\epsfxsize=7cm \epsfbox{ptlhc.120}
\vspace*{-9.75cm}

\hspace*{8.5cm}
\epsfxsize=7cm \epsfbox{ptlhc.500}
\vspace*{-1.5cm}

\caption[]{\label{fg:bbh} \it Transverse momentum distributions of the
bottom quark in $b\bar bH$ production for two different Higgs masses at
the LHC. We have adopted CTEQ5M1 parton densities and a bottom mass of
$m_b=4.62$ GeV. The solid lines show the full LO result from
$q\bar q,gg\to b\bar bH$ and the dashed lines the factorized collinear
part, which is absorbed in the bottom parton density. The upper curves
are divided by the factor $p_{Tb}/(m_b^2+p_{Tb}^2)$ of the asymptotic
behavior, which is required by factorizing bottom densities.}
\end{figure} 
The transverse momentum distributions at the LHC
are shown in Fig.~\ref{fg:bbh}, for two different Higgs
masses. The solid curves show the full distributions of the $q\bar
q,gg\to b\bar b\phi$ processes, while the dashed lines exhibit the factorized
collinear part, which is absorbed in the bottom density. For a proper
factorization, these pairs of curves have to coincide approximately up to
transverse momenta of the order of the factorization scale, which is
usually chosen to be $\mu_F={\cal O}(m_H)$. It is clearly visible that
there are sizeable differences between the full result and the
factorized part, which originate from sizeable bottom mass and phase
space effects, that are not accounted for by an active bottom
parton density. Moreover, the full result falls quickly
below the approximate factorized part for transverse momenta of the
order of $m_H/10$, which is much smaller than the usual factorization
scale used for the bottom densities. We conclude from these plots that
$b\bar b\phi$ production at the LHC develops sizeable bottom mass effects,
so that the use of bottom densities in the process $b\bar b\to \phi$ may
lead to an overestimate of the correct theoretical result due to too
crude approximations in the kinematics of the hard process. The full NLO
calculation of the $gg\to b\bar b\phi$ will
yield much more insight into this problem, since the large logarithms
related to the evolution of bottom densities have to appear in the
NLO corrections, if the picture of active bottom quarks in the proton is
correct.

\section{$ZH,WH$ production}
Higgsstrahlung in $q\bar q\to WH,ZH$ plays a crucial role for the Higgs
search at the Tevatron, while it is only marginal at the LHC. At the
Tevatron it provides the relevant production mechanism for Higgs masses
below about 130 GeV, where $H\to b\bar b$ decays are dominant~\cite{run2}. 
The NLO QCD corrections have been analyzed in the past. They are identical 
to the QCD corrections to the Drell--Yan processes $q\bar q\to W,Z$, if the
LO matrix elements are replaced accordingly. QCD corrections increase
the production cross sections by about 30--40\%~\cite{vhnlo,susyqcd}.

The most important backgrounds at the Tevatron are $Wjj$ and in particular
$Wb\bar b$ production. Both are known at NLO and are contained in a NLO
Monte Carlo program~\cite{MCFM}.  The
same applies also to the $Zjj$ and in particular $Zb\bar b$ 
backgrounds~\cite{veseli,zjjNLO}.
In addition, the $t\bar t$ background is relevant.

\section{Conclusions}
Considerable progress has been made recently in improving QCD calculations
for Higgs signal and background cross sections at hadron colliders.
Noteworthy examples are the NNLO corrections to the gluon fusion cross 
section~\cite{Harlander:2002wh}, the QCD $Zjj$ cross section at 
NLO~\cite{zjjNLO} and the determination of full finite top and $W$ width 
corrections to $t\bar t$ and $t\bar tj$ production at LO~\cite{ttoff}.
These improvements are crucial for precise coupling determinations of the
Higgs boson.

Much additional work is needed to match the statistical power of the LHC.
Largely, QCD systematic errors for coupling measurements have not been 
analyzed yet. Additional NLO tools need to be provided as
well, and these include NLO corrections to $t\bar t$ production with 
finite width effects and $t\bar tj$ production at zero top width.
A better understanding of central jet veto efficiencies is crucial for 
the study of WBF channels. These are just a few examples where theoretical 
work is needed. Many more have been highlighted in this review. 
Higgs physics at the LHC remains a very rich field 
for phenomenology. 

\vspace{0.2in}
\noindent {\bf Acknowledgments.} \\
We would like to thank the organizers of the Les Houches workshop for
their invitation, warm hospitality and financial support. The work of M.S.
has been supported in part by the
Swiss Bundesamt f\"ur Bildung und Wissenschaft and by the European Union
under contract HPRN-CT-2000-00149.
The work of D.Z. was partially supported by WARF and under DOE grant 
No.~DE-FG02-95ER40896. 
Fermilab is operated by URA under DOE contract No.~DE-AC02-76CH03000.


\begin{thebibliography}{99}

\bibitem{higgs} P.\ W.\ Higgs,
Phys.\ Lett.\ {\bf 12} (1964) 132 and
Phys.\ Rev.\ {\bf 145} (1966) 1156; \\
F.\ Englert and R.\ Brout, Phys.\ Rev.\ Lett.\ {\bf 13} (1964) 321; \\
G.\ S. Guralnik, C.\ R.\ Hagen and T.\ W.\ Kibble, Phys.\ Rev.\ Lett.\
{\bf 13} (1964) 585.

\bibitem{hunter} For reviews
on the Higgs sector in the Standard Model and in
its supersymmetric extensions, see J.F.\ Gunion, H.E.\ Haber, G.L.\ Kane
and S.\ Dawson, {\it The Higgs Hunter's Guide} (Addison--Wesley, Reading,
Mass., 1990).

\bibitem{lep2}
LEP Higgs Working Group for Higgs boson searches,
Proceedings International Europhysics Conference on High Energy
Physics (HEP 2001), Budapest, Hungary, 12-18 Jul 2001, hep-ex/0107029
and hep-ex/0107030.

\bibitem{triviality}
See e.g.\,T.\,Hambye and K.\,Riesselmann, Phys.\,Rev.\,{\bf D55} (1997)
7255.

\bibitem{mssmrad}
M.\ Carena, M.\ Quiros and C.E.M.\ Wagner, Nucl.\ Phys.\ {\bf B461}
(1996) 407;
H.E.\ Haber, R.\ Hempfling and A.H.\ Hoang, Z.\ Phys.\ {\bf C75} (1997)
539;
S.\ Heinemeyer, W.\ Hollik and G.\ Weiglein, Phys.\ Rev.\ {\bf D58}
(1998) 091701;
R.--J.~Zhang, Phys.~Lett.~{\bf B447} (1999) 89;
J.\,Espinosa and R.--J.\,Zhang, Nucl.\,Phys.\,{\bf B586} (2000) 3;
A.\,Brignole, G.\,Degrassi, P.\,Slavich and F.\,Zwirner, hep-ph/0112177.

\bibitem{run2} M.~Carena et al., Proceedings `Physics at Run II: Workshop
on Supersymmetry/Higgs', Batavia, IL, 19-21 Nov 1998, hep-ph/0010338.
 
\bibitem{lhc}
ATLAS Collaboration, Technical Design Report, CERN--LHCC 99--14 (May 1999);
CMS Collaboration, Technical Proposal, CERN--LHCC 94--38 (Dec.~1994).
 
\bibitem{Zeppenfeld:2000td}
D.~Zeppenfeld, R.~Kinnunen, A.~Nikitenko and E.~Richter-Was,
Phys.\ Rev.\ {\bf D62} (2000) 013009.

\bibitem{difractH}
A.~Bialas and P.~V.~Landshoff,
Phys.\ Lett.\ {\bf B256} (1991) 540;
J.-R. Cudell and O.F. Hernandez, Nucl. Phys. {\bf B471} (1996) 471; 
V.~A.~Khoze, A.~D.~Martin and M.~G.~Ryskin,
hep-ph/0111078 and references therein.

\bibitem{habil}
M.~Spira, Fortschr.~Phys.~{\bf 46} (1998) 203.

\bibitem{squark}
S.\ Dawson, A.\ Djouadi and M.\ Spira, Phys.\ Rev.\ Lett.\ {\bf 77} (1996) 16.

\bibitem{glufusnlo}
A.\ Djouadi, M.\ Spira and P.M.\ Zerwas, Phys.\ Lett.\ {\bf B264}
(1991) 440;
S.\ Dawson, Nucl.\ Phys.\ {\bf B359} (1991) 283;
D.\ Graudenz, M.\ Spira and P.M.\ Zerwas, Phys.\ Rev.\ Lett.\
{\bf 70} (1993) 1372;
S.\ Dawson and R.P.\ Kauffman, Phys.\ Rev.\ {\bf D49} (1994)
2298;
M.\ Spira, A.\ Djouadi, D.\ Graudenz and P.M.\ Zerwas, Phys.\
Lett.\ {\bf B318} (1993) 347,
Nucl.\ Phys.\ {\bf B453} (1995) 17.
 
\bibitem{3loop}
R.V.~Harlander, Phys.~Lett.~{\bf B492} (2000) 74.
 
\bibitem{Harlander:2002wh}
R.~V.~Harlander and W.~B.~Kilgore,
hep-ph/0201206.

\bibitem{softgluon}
M.~Kr\"amer, E.~Laenen and M.~Spira, Nucl.~Phys.~{\bf B511} (1998) 523.
 
\bibitem{soft}
S.~Catani, D.~de Florian and M.~Grazzini, JHEP {\bf 0105} (2001) 025;
R.V.~Harlander and W.B.~Kilgore, Phys.~Rev.~{\bf D64} (2001) 013015.
 
\bibitem{soft2} S.\,Catani, D.\,de Florian, M.\,Grazzini and P.\,Nason,
to appear in the 
proceedings of the 2001 Les Houches workshop on ``Physics at TeV Colliders''.
 
\bibitem{drollinger:2001ym}
V.~Drollinger, T.~Muller and D.~Denegri,
hep-ph/0111312.

\bibitem{diphox}
T.~Binoth, J.P.~Guillet, E.~Pilon and M.~Werlen, Eur.~Phys.~J.~{\bf C16}
(2000) 311.

\bibitem{gggamgam}
Z.\,Bern, A.\,De Freitas and L.J.\,Dixon,  JHEP {\bf 09} (2001) 037.
 

\bibitem{Catani:2001cr}
S.~Catani, D.~de Florian and M.~Grazzini,
JHEP {\bf 0201} (2002) 015.


\bibitem{vvnlo}
S.~Frixione, P.~Nason and G.~Ridolfi, Nucl.~Phys.~{\bf B383} (1992) 3;
S.~Frixione, Nucl.~Phys.~{\bf B410} (1993) 280;
U.~Baur, T.~Han and J.~Ohnemus, Phys.~Rev.~{\bf D48} (1993) 5140,
Phys.~Rev.~{\bf D51} (1995) 3381, Phys.~Rev.~{\bf D48} (1996) 1098,
Phys.~Rev.~{\bf D57} (1998) 2823.

\bibitem{kunszt}
L.J.\,Dixon, Z.\,Kunszt and A.\,Signer, Phys.\,Rev.\,{\bf D60} (1999)
114037.

\bibitem{offshell}
T.G.~Rizzo, Phys.~Rev.~{\bf D22} (1980) 389;
W.-Y.~Keung and W.J.~Marciano, Phys.~Rev.~{\bf D30} (1984) 248;
R.N.\ Cahn, Rep.\ Prog.\ Phys.\ {\bf 52} (1989) 389.

\bibitem{dittdrei}
M.\ Dittmar and H.\ Dreiner, Phys.\ Rev.\ {\bf D55} (1997) 167.

\bibitem{ttoff}
N.~Kauer and D.~Zeppenfeld,
Phys.\ Rev.\ {\bf D65} (2002) 014021.

\bibitem{veseli}
R.K.~Ellis and S.~Veseli, Phys.~Rev.~{\bf D60} (1999) 011501 and
Nucl.~Phys.~{\bf B511} (1998) 649;
J.M.~Campbell and R.K.~Ellis, Phys.~Rev.~{\bf D62} (2000) 114012.

\bibitem{vvhnlo}
T.\ Han, G.\ Valencia and S.\ Willenbrock, Phys.\ Rev.\ Lett.\
{\bf 69} (1992) 3274.

\bibitem{susyqcd} A.\,Djouadi and M.\,Spira, Phys.\,Rev.\,{\bf D62}
(2000) 014004.

\bibitem{wbfgamma}
D.~Rainwater and D.~Zeppenfeld,
JHEP {\bf 9712} (1997) 005.

\bibitem{wbftau}
D.~Rainwater, D.~Zeppenfeld and K.~Hagiwara,
Phys.\ Rev.\ {\bf D59} (1999) 014037;
T.~Plehn, D.~Rainwater and D.~Zeppenfeld,
Phys.\ Rev.\ {\bf D61} (2000) 093005.

\bibitem{wwhimH}
D.~Rainwater and D.~Zeppenfeld,
Phys.\ Rev.\ {\bf D60} (1999) 113004,
(E) {\it ibid.}\ {\bf D61} (1999) 099901.

\bibitem{wbf.wwlomh}
N.~Kauer, T.~Plehn, D.~Rainwater and D.~Zeppenfeld,
Phys.\ Lett.\ {\bf B503} (2001) 113.

\bibitem{vvhtau}
T.~Plehn, D.~Rainwater and D.~Zeppenfeld,
Phys.\ Lett.\ {\bf B454} (1999) 297.

\bibitem{hjjmass}
V.~Del Duca, W.~Kilgore, C.~Oleari, C.~Schmidt and D.~Zeppenfeld,
Nucl.\ Phys.\ {\bf B616} (2001) 367.

\bibitem{wbf.exp}  
G.~Azuelos et al., {\it Search for the Standard Model Higgs Boson 
using Vector Boson Fusion at the LHC}, in D.~Cavalli et al., 
``The Higgs working group: Summary report'' for Les Houches 2001,
arXiv:hep-ph/0203056.

\bibitem{hjjlimit}
S.~Dawson, R.P.~Kauffman, Phys.\ Rev.\ Lett.\ {\bf 68} (1992) 2273;
R.P.~Kauffman, S.V.~Desai, D.~Risal, Phys.\ Rev.\ {\bf D55} (1997) 4005,
(E) {\it ibid.}\ {\bf D58} (1998) 119901,
Phys.\ Rev.\ {\bf D59} (1999) 057504. 

\bibitem{zjjNLO}
J.~Campbell and R.~K.~Ellis,
hep-ph/0202176.

\bibitem{tauola}
S.~Jadach, J.H.~K\"uhn and Z.~Was, Comput.~Phys.~Commun.~{\bf 64} (1990) 275; 
S.~Jadach, Z.~Was, R.~Decker and J.H.~K\"uhn, Comput.~Phys.~Commun.~{\bf 76}
(1993) 361.

\bibitem{Chehime:1992ub}
H.~Chehime and D.~Zeppenfeld,
Phys.\ Rev.\ {\bf D47} (1993) 3898;
D.~Rainwater, R.~Szalapski and D.~Zeppenfeld,
Phys.\ Rev.\ {\bf D54} (1996) 6680.

\bibitem{MCFM}
J.\,Campbell and R.K.\,Ellis, 
http://theory.fnal.gov/people/ellis/Programs/mcfm.html

\bibitem{tth2bb.atlas}
M.~Sapinski and D.~Cavalli,
Acta Phys.\ Polon.\ {\bf B32} (2001) 1317;
E.~Richter-Was and M.~Sapinski,
Acta Phys.\ Polon.\ {\bf B30} (1999) 1001.

\bibitem{tth2bb.cms}
D.~Green, K.~Maeshima, R.~Vidal, W.~Wu and S.~Kunori,
FERMILAB-FN-0705.

\bibitem{tth2ww}
F.~Maltoni, D.~Rainwater and S.~Willenbrock,
hep-ph/0202205.

\bibitem{tthnlo}
W.~Beenakker, S.~Dittmaier, M.~Kr\"amer, B.~Pl\"umper, M.~Spira and
P.M.~Zerwas, Phys.\,Rev.\,Lett.\,{\bf 87} (2001) 201805.

\bibitem{tthnloq}
L.~Reina and S.~Dawson, Phys.\,Rev.\,Lett.\,{\bf 87} (2001) 201804.

\bibitem{willenbrock}
D.A.\ Dicus and S.\ Willenbrock, Phys.\ Rev.\ {\bf D39} (1989) 751.

\bibitem{uninpdf}
J.\, Kwiecinski, A.D.\, Martin and A.M.\, Stasto, Phys.\, Rev.\, {\bf D56}
(1997) 3991;
M.A.\, Kimber, A.D.\, Martin and M.G.\, Ryskin, Eur.\, Phys.\, J.\, {\bf C12}
(2000) 655.

\bibitem{vhnlo}
T.\ Han and S.\ Willenbrock, Phys.\ Lett.\ {\bf B273} (1991) 167.

\end{thebibliography}
\end{document}